\documentclass[aps,pra,showpacs,showkeys,twocolumn,twoside]{revtex4}

\usepackage{amsfonts}
\usepackage{amssymb}
\usepackage{amsmath}
\usepackage{amscd}
\usepackage{latexsym}  % maybe this one is needed, too.
\usepackage{epsfig}
\usepackage{bbm}

\newtheorem{defi}{Definition}

\newtheorem{satz}[defi]{Theorem}

\newtheorem{rem}[defi]{Remark}

\newtheorem{exempel}[defi]{Example}

\newcommand{\qed}{\hfill $\Box$}
\newcommand{\tr}{{\operatorname{Tr}\,}}
\newcommand{\id}{{\operatorname{id}}}

\newcommand{\bra}[1]{{\langle{#1}|}}
\newcommand{\ket}[1]{{|{#1}\rangle}}
\newcommand{\ketbra}[1]{{\ket{#1}\!\bra{#1}}}
\newcommand{\C}{{\mathbb{C}}}

\newcommand{\N}{{\mathbb{N}}}
\newcommand{\alg}[1]{{\mathfrak{#1}}}

\newcommand{\1}{{\openone}}

\newlength{\blank}
\newlength{\equalsign}
\newenvironment{beweis}[1][{\hspace{-\blank}}]{{\noindent\emph{Proof~{#1}.\ }}}{\hfill $\Box$\vskip 0.5\baselineskip}

\begin{document}

\title{Scalable programmable quantum gates and\\ a new aspect of the additivity problem for\\ the classical capacity of quantum channels}
\author{Andreas Winter}
\email{winter@cs.bris.ac.uk}
\affiliation{Department of Computer Science, University of Bristol, Merchant Venturers Building, Woodland Road, Bristol BS8 1UB, United Kingdom.}
\date{October 22, 2001}

%begonnen: 25. Juli 2001

% abstract

\begin{abstract}
  We consider two apparently separated problems:
  in the first part of the paper we study the concept
  of a \emph{scalable (approximate) programmable quantum gate} (SPQG).
  These are special (approximate) programmable quantum gates,
  with nice properties that could have implications on the theory
  of universal computation. Unfortunately, as we prove, such objects
  do not exist in the domain of usual quantum theory.
  \par
  In the second part the problem of noisy dense coding
  (and generalizations) is addressed. We observe that the additivity
  problem for the classical capacity obtained is
  of apparently greater generality
  than for the usual quantum channel (completely positive maps):
  i.e., the latter occurs as a
  special case of the former, but, as we shall argue with the help
  of the non--existence result of the first part, the former cannot
  be reduced to an instance of the latter.
  \par
  We conclude by suggesting that the additivity problem for the
  classical capacity of quantum channels, as posed until now, may
  conceptually not be in its appropriate generality.
\end{abstract}



\maketitle

% hier gehts los...

\section{Introduction}
\label{sec:intro}
The present paper brings together two subjects in the realm of quantum
information theory that might at first glance seem far apart: the theory
of universal computation in a quantum computer, and noise resistant
coding of classical information in quantum channels.
\par
The former deals with implementing arbitrary transformation of the
(quantum) data in the memory of a computer by a sequence of commands
(a \emph{program}) that are themselves presented to the machine as data.
From the first days of the theory of quantum computation this issue
was of central importance, as a tool to show that there is essentially
only one quantum Turing machine, and to parallel Turing's
insight of the existence of universal classical machines:
see Deutsch~\cite{deutsch:universal}, and Bernstein and
Vazirani~\cite{bv:universal}. A great deal of work has been invested
into finding small universal sets of ``quantum gates'', acting on only
few qubits at a time, so that by concatenation any multi--qubit
unitary can be approximated arbitrarily (Deutsch~\cite{deutsch:gate},
DiVincenzo~\cite{divincenzo:gate}, Barenco~\cite{barenco:gates},
Deutsch et al.~\cite{dbe:gates}, and Barenco et al.~\cite{bbcdmsssw:gates}).
This concatenation (represented as a certain directed graph
with labelled nodes) can be given to a
machine as classical data, which then interpretes it as a
series of controlled actions on the quantum data.
\par
The universality problem was studied abstractly
by Nielsen and Chuang~\cite{nielsen:chuang}
in the notion of \emph{programmable quantum gate} (PQG), where
one allows \emph{arbitrary quantum} data for a program,
their results being further developed by Vidal, Masanes, and
Cirac~\cite{vidal:cirac,vidal:masanes:cirac}. We review these studies,
as far as they are relevant for the present
purpose, in section~\ref{sec:pqg}.
Then, in section~\ref{sec:spqg}, the notion of \emph{scalability}
is introduced, which captures the idea that a sufficiently powerful
universal programmable quantum gate might give a universal gate
if tensored with itself and fed with entangled programs.
Unfortunately, it turns out that such objects do not exist,
and we point out some implications for the general theory of
universal computation.
\par
Then, we switch to the apparently completely distinct problem
of quantum channel coding of classical data: a quantum channel
usually is modelled by a completely positive, trace preserving,
linear map $\varphi:{\cal B}({\cal H}_1)\rightarrow{\cal B}({\cal H}_2)$.
We may use this channel to communicate by choosing states
$\sigma_i$ on ${\cal H}_1$ at the sender's side, the receiver
getting $\varphi(\sigma_i)$. By the result of Holevo~\cite{holevo:coding},
and Schumacher and Westmoreland~\cite{sw:coding} the maximum rate
at which classical information can be transmitted asymptotically reliably
(the \emph{capacity}) is given by
\begin{equation}
  \label{eq:capacity}
  C(\varphi)=\max_{\{\sigma_i,p_i\}} I(p;\varphi(\sigma)),
\end{equation}
with the \emph{Holevo mutual information}
$$I(p;\varphi(\sigma))=H\left(\sum_i p_i\varphi(\sigma_i)\right)
                       -\sum_i p_i H(\varphi(\sigma_i)),$$
$H$ being the von Neumann entropy on density operators.
This holds when in the block coding (implicit in the statement)
for $\varphi^{\otimes n}$ one is restricted to using product
states $\sigma_{i_1}\otimes\ldots\otimes\sigma_{i_n}$.
Strictly speaking, we should write ``$\sup$'' over \emph{all}
probability distributions $p$ on states, and integrals instead of
finite sums. However, we restrict our attention to finite dimensional
spaces, and there it is possible to show that the supremum is achieved
by a finitely supported measure $p$ (see~\cite{sw:coding}).
\par
Unfortunately it is unknown whether it is sufficient to restrict
coding to product states. It would be if the additivity conjecture
$$C(\varphi\otimes\vartheta)=C(\varphi)+C(\vartheta)$$
is true. To show this one would need to consider input state
ensembles with entangled states, and prove that the corresponding
Holevo information can be achieved by an ensemble without entangled
states, or more directly, that a code using entangled states can
be modified to an equally good code (in terms of error probability
and rate) without entangled states. Neither of these has been achieved
in generality so far, though there have been advances
recently: see~\cite{ahw:add} and~\cite{king:add}.
\par
In section~\ref{sec:noisy:dense:coding} we present an example
of a special classical--quantum channel as a case study:
dense coding in the presence of noisy entanglement, and by
use of a general quantum channel, in particular a noiseless
one. Here, coding is done by selecting not a state of a system,
to be sent down the channel, but by selecting an operation
on a given state. This is a more general concept of coding,
as we demonstrate in section~\ref{sec:reductions}.
It appears that the coding of such a channel
can be approximated by programmable quantum gates (in this sense
the new model is a special case of the old one), but that
the parallel use of these systems cannot: there will
always be actions on the combined space that cannot be mimicked
by entangled inputs to the PQG--augmented channel.
\par
We conclude with the suggestion that the additivity problem
for classical capacities of quantum channels has not been
posed until now in its appropriate generality.

\section{Programmable quantum gates}
\label{sec:pqg}
In classical computers there is no fundamental distinction in a
univeral machine's memory between data and program. In fact a program
may modify itself during the computation (a feature considered essential
by von Neumann when he designed his computer model).
To which extent can a quantum computer memory be used to modify
other parts of the memory in a program--like fashion?
More precisely (following~\cite{nielsen:chuang}): assume that a unitary
process $G$ acts on ${\cal H}_D\otimes{\cal H}_P$, with
the \emph{data register} ${\cal H}_D$ and the \emph{program register}
${\cal H}_P$:
$$\ket{\zeta}\otimes\ket{\psi}\longmapsto
                         G\left(\ket{\zeta}\otimes\ket{\psi}\right).$$
We call $\ket{\psi}$ a \emph{program} if it has the property that
\begin{equation}
  \label{eq:program}
  \forall\ket{\zeta}\quad G\left(\ket{\zeta}\otimes\ket{\psi}\right)
                                   =U_\psi\ket{\zeta}\otimes\ket{\psi'}.
\end{equation}
Note that --- though a priori $\ket{\psi'}$ could also depend on $\ket{\zeta}$
--- for $\ket{\zeta_1},\ket{\zeta_2}\in{\cal H}_D$ the corresponding
$\ket{\psi_1'},\ket{\psi_2'}$ are linearly dependent:
\begin{equation*}\begin{split}
  G &\bigl( (\alpha\ket{\zeta_1}+\beta\ket{\zeta_2})\otimes\ket{\psi} \bigr) \\
    &\phantom{=====}= \alpha U_\psi\ket{\zeta_1}\otimes\ket{\psi_1'}
                     +\beta U_\psi\ket{\zeta_2}\otimes\ket{\psi_2'},
\end{split}\end{equation*}
which generally is entangled unless $\ket{\psi_1'}\in\C\ket{\psi_2'}$.
\par
(We thus \emph{can} have a global phase --- which we shall systematically
ignore). Henceforth we assume that $\ket{\psi'}$ is
independent of $\ket{\zeta}$, just as equation~\ref{eq:program} suggests.
It follows that $U_\psi$ is unitary, which is encoded (via $G$)
in the program $\ket{\psi}$.
How many unitaries can be implemented in this way?
\par
\begin{satz}[Nielsen,~Chuang~\cite{nielsen:chuang}]
  \label{satz:no:upqg}
  If $U_{\psi_1}\neq\gamma U_{\psi_2}$ for all $\gamma\in\C$,
  then $\ket{\psi_1}\perp\ket{\psi_2}$.
\end{satz}
\begin{beweis}
  Let
  \begin{align*}
    G\left(\ket{\zeta}\otimes\ket{\psi_1}\right)
                              &=U_{\psi_1}\ket{\zeta}\otimes\ket{\psi_1'},   \\
    G\left(\ket{\zeta}\otimes\ket{\psi_2}\right)
                              &=U_{\psi_2}\ket{\zeta}\otimes\ket{\psi_2'}.
  \end{align*}
  Hence
  \begin{align*}
    \langle\psi_1\ket{\psi_2} &=\left(\bra{\zeta}\otimes\bra{\psi_1}\right)G^*
                                G\left(\ket{\zeta}\otimes\ket{\psi_2}\right) \\
                              &=\langle\psi_1'\ket{\psi_2'}
                                \bra{\zeta}U_{\psi_1}^* U_{\psi_2}\ket{\zeta}.
  \end{align*}
  If $\langle\psi_1'\ket{\psi_2'}=0$, also $\langle\psi_1\ket{\psi_2}=0$,
  and we are done. Else $\bra{\zeta}U_{\psi_1}^* U_{\psi_2}\ket{\zeta}$
  is a constant, independent of $\ket{\zeta}$, hence
  $U_{\psi_1}^* U_{\psi_2}=\gamma\1$, contradicting the assumption.
\end{beweis}
As a consequence we have only at most $\log\dim{\cal H}_P$
many essentially different programs. There is no way to encode
all possible unitaries on ${\cal H}_D$ by ``quantum code''
unless we allow for an infinite--dimensional ${\cal H}_P$.
\par\medskip
We have already in the introduction pointed out that it is well possible
to implement arbitarily good approximations to all unitaries (at the cost
of ever increasing $\dim{\cal H}_P$).
In~\cite{nielsen:chuang}, however, there was proposed a more interesting
solution: a \emph{probabilistic} programmable quantum gate,
i.e. an encoding of unitaries in a state, and a process that performs the
encoded unitary with some probability, and otherwise fails (does something
else): the process is able to report which of the two events happened.
This result was refined in subsequent work of
Vidal, Masanes, and Cirac~\cite{vidal:cirac,vidal:masanes:cirac},
but we will not follow this line of research here.
\par\medskip
To fix notions, let us define our concept of approximation:
a (unitary) gate $G$ on ${\cal H}_D\otimes{\cal H}_P$
is said to be \emph{$\epsilon$--approximating} if for 
every unitary $U$ on ${\cal H}_D$ there is a state vector
$\ket{\psi}\in{\cal H}_P$ (it is easily seen that pure state program
register contents suffice) such that
$$\forall\ket{\zeta}\quad \left\| U\ketbra{\zeta}U^*
                                   - \tr_{{\cal H}_P} \bigl(G(\ketbra{\zeta}\otimes\ketbra{\psi})G^*\bigr)
                          \right\|_1\leq \epsilon.$$
Of course there are $\epsilon$--approximating gates such that
the approximating induced maps
$$\Gamma_\psi(\sigma)=\tr_{{\cal H}_P} \bigl(G(\sigma\otimes\ketbra{\psi})G^*\bigr)$$
in the above equation all may be chosen unitary, but the present formulation
has the appropriate generality for the nonexistence theorem of the
following section.
\par
A sequence $(G^{(n)})_{n\in\N}$ of programmable quantum gates
$G^{(n)}$ on ${\cal H}_{P_n}\otimes{\cal H}_D$ is called
\emph{approximating for ${\cal H}_D$} if each $G^{(n)}$ is $\epsilon_n$--approximating,
with $\epsilon_n\rightarrow 0$ for $n\rightarrow\infty$.

\section{Scalable programmable\protect\\ quantum gates}
\label{sec:spqg}
Given $\epsilon>0$ we can devise $\epsilon$--approximating quantum gates
$G_1$ and $G_2$ for given data registers ${\cal H}_{D_1}$ and ${\cal H}_{D_2}$,
respectively, by allowing for sufficiently large program registers.
\par
Programming, however, is about making act together data in a potentially
unlimited number of registers. In general, to approximately
perform an arbitary unitary on ${\cal H}_{D_1}\otimes{\cal H}_{D_2}$
it is necessary to define a new quantum gate $G$.
\par
This motivates us to the following definition:
we say that two sequences $(G^{(n)}_1)_{n\in\N}$ and $(G^{(n)}_2)_{n\in\N}$
of programmable quantum gates that are approximating for ${\cal H}_{D_1}$
and ${\cal H}_{D_2}$, respectively, are \emph{scalable}, if the
sequence $(G^{(n)}_1\otimes G^{(n)}_2)_{n\in\N}$ is approximating
for ${\cal H}_{D_1}\otimes{\cal H}_{D_2}$.
\par
Such approximating gate sequences thus spare us the task to find
an implement new programmable quantum gates when we scale up
our computing system.
\par
Unfortunately, nature does not supply us with such objects:
\begin{satz}
  \label{satz:no:spqg}
  Let $(G^{(n)}_1)_{n\in\N}$ and $(G^{(n)}_2)_{n\in\N}$ be sequences of
  programmable quantum gates with fixed data registers
  ${\cal H}_{D_1}$ and ${\cal H}_{D_2}$, respectively. Assume that
  the unitary $U$ on ${\cal H}_{D_1}\otimes{\cal H}_{D_2}$
  is approximated arbitarily close by programs
  $\psi^{(n)}\in {\cal H}_{P_{1 n}}\otimes{\cal H}_{P_{2 n}}$, i.e.
  \begin{equation}
    \label{eq:spqg}
    \begin{split}
    \tr&_{{\cal H}_{P_{1 n}}\otimes{\cal H}_{P_{2 n}}} \!
          \left[  G^{(n)}_1\!\otimes G^{(n)}_2
               \bigl(\ketbra{\zeta}\otimes\ketbra{\psi}\bigr)
          G^{(n)*}_1\!\otimes G^{(n)*}_2 \right]           \\
       &\phantom{=}\longrightarrow U\ketbra{\zeta}U^*
  \end{split}\end{equation}
  as $n\rightarrow\infty$.
  Then $U$ is not entangling, i.e. it is of the form $U=U_1\otimes U_2$.
\end{satz}
\begin{beweis}
  Consider the expressions of eq.~(\ref{eq:spqg}) for data of the
  form $\ket{\zeta}=\ket{\zeta_1}\otimes\ket{\zeta_2}$. The first
  claim is that the reduced state of the left hand side on
  ${\cal H}_{D_1}$ is independent of $\zeta_2$: this becomes clear
  by first tracing out ${\cal H}_{D_2}\otimes{\cal H}_{P_{2 n}}$
  and then ${\cal H}_{P_{2 n}}$.
  Then the same applies to the limit at the right hand side.
  \par
  So, for fixed $\ket{\zeta_1}$ we have
  \begin{equation}
    \label{eq:mixed}
    \begin{split}
    \tr_{{\cal H}_{D_2}} U\bigl(\ketbra{\zeta_1}\otimes\ketbra{\zeta_2}\bigr)U^*
                     &= \rho_0 \\
                     &= \sum_i \lambda_i\ketbra{e_i},
    \end{split}
  \end{equation}
  with a constant state $\rho_0$ (that we wrote in diagonalized form),
  regardless of $\ket{\zeta_2}$.
  \par
  Now assume that $U$ is entangling,
  and choose $\ket{\zeta_1}$ such there exists $\ket{\zeta_2}$
  so that $U\ket{\zeta_1}\otimes\ket{\zeta_2}$ is entangled.
  Then $\rho_0$ is mixed, and its diagonalization contains
  at least two terms. We shall derive a contradiction from
  this:
  first observe that for arbitrary $\ket{\zeta_2}$
  the state $U\ket{\zeta_1}\otimes\ket{\zeta_2}$
  is a purification of $\rho_0$, hence, by eq.~(\ref{eq:mixed}) there exists
  an orthonormal basis $\{\ket{f_i}\}$ of ${\cal H}_{D_2}$ such that
  $$U\ket{\zeta_1}\otimes\ket{\zeta_2}=\sum_i \sqrt{\lambda_i}\ket{e_i}\otimes\ket{f_i}.$$
  For $\ket{\zeta_2'}$ orthogonal to $\ket{\zeta_2}$ there is another
  such basis $\{\ket{f_i'}\}$ with 
  $$U\ket{\zeta_1}\otimes\ket{\zeta_2'}=\sum_i \sqrt{\lambda_i}\ket{e_i}\otimes\ket{f_i'}.$$
  By linearity we get thus
  $$U\ket{\zeta_1}\otimes(\alpha\ket{\zeta_2}+\beta\ket{\zeta_2'})
                =\sum_i \sqrt{\lambda_i}\ket{e_i}\otimes(\alpha\ket{f_i}+\beta\ket{f_i'}),$$
  for $|\alpha|^2+|\beta|^2=1$. This again must be a purification of $\rho_0$,
  so the resulting $\{\alpha\ket{f_i}+\beta\ket{f_i'}\}$
  must form an orthonormal basis:
  this leads quickly to the condition (for all $i,j$)
  $$\overline{\alpha}\beta\langle f_i\ket{f_j'}
      +\alpha\overline{\beta}\langle f_i'\ket{f_j}=0,$$
  implying $\langle f_i\ket{f_j'}=\langle f_i'\ket{f_j}=0$, otherwise
  $z$ and $\overline{z}$ would be linearly dependent over the complex field.
  \par
  As a consequence, to each orthonormal system of $\ket{\zeta_2}$'s of
  ${\cal H}_{D_2}$ we would get an orthonormal system of $\ket{f_i}$'s
  of at least double size, contradicting the finite dimension of
  ${\cal H}_{D_2}$. Thus $U$ cannot be entangling, forcing
  $U=U_1\otimes U_2$. To see this either consult~\cite{durt}
  or follow this simple argument: since
  $$\sigma_{12}\longmapsto U\sigma_{12}U^*$$
  mpas product states to product states, the map
  $$T_1:\sigma\longmapsto \tr_{2}U\bigl(\sigma\otimes\ketbra{\zeta_2}\bigr)U^*$$
  maps pure states to pure states and is completely positive,
  entailing that it has to be of the form
  $T_1(\sigma)=U_1\sigma U_1^*$ (this may be viewed as an easy kind
  of Wigner--theorem). Here $U_1$ is a unitary which cannot --- except
  for a global phase --- depend of $\ket{\zeta_2}$, or else there
  would be entangled states $U(\sigma_1\otimes\sigma_2)U^*$.
  The same applies to the second factor, yielding a unitary $U_2$.
  In total we have that the unitary $U_1\otimes U_2$ coincides with
  $U$ on te pure states, hence $U=U_1\otimes U_2$
  (again except for an unimportant global phase).
\end{beweis}
Observe the following peculiarity of the argument: it is not true that
the reduced state at the left hand side of eq.~(\ref{eq:spqg}) is always a product
(if it is, our proof is simplified drastically). For example $G_1$ and $G_2$
may be swapping operations, so their product may be used to swap in
any entangled state! What is true however is, that entangled states
cannot occur as a result of a unitary action on the data registers.
\par
This nonexistence should not be mixed up with the existence of the beautiful
model of \emph{one--way quantum computer} by Raussendorf and
Briegel~\cite{raussen:briegel}: there, too, a single state is prepared
and acted on locally (even only by measurements), to produce any
given effect on the data register. There is no coantradiction, however,
to our result, as there is implied \emph{classical} communication
between the sites of these quantum operations, which we had to exclude.
\par\medskip
In a sense, the result had to be expected: it reproduces on a somewhat different
level the insight in universal computation that single qubit actions are not
sufficient for universality, but one needs interacting gates like the
C--NOT gate.
\par
We shall show in the following, however, that this nonexistence result has some
bearing on quantum channel coding.

\section{Noisy dense coding capacity}
\label{sec:noisy:dense:coding}
Consider the following communication scenario:
a sender A and a receiver B share a state $\rho$ on
the $d_A\times d_B$--system ${\cal H}_A\otimes{\cal H}_B$, i.e.
$\dim{\cal H}_A=d_A$, $\dim{\cal H}_B=d_B$. They have at their disposal
a quantum channel from A to B that allows noiseless
transmission of an arbitrary quantum state in ${\cal H}\simeq\C^{d}$.
They want to use this channel to communicate classical information,
taking advantage of the correlation (or even entanglement) of
$\rho$. The most general thing possible for A to do is to
subject her share of the state to an operation, and send the result
through the channel.
It is well known that, if $\rho$ supplies only
classical correlation (for instance, if
$$\rho=\sum_{i=0}^{d_A-1}\sum_{j=0}^{d_B-1} p_{ij} \ketbra{i}\otimes\ketbra{j},$$
for orthogonal bases $\{\ket{i}:i=0,\ldots,d_A-1\}$ and
$\{\ket{j}:j=0,\ldots,d_B-1\}$ of ${\cal H}_A$ and
${\cal H}_B$, respectively), then this is of no help at all,
and the capacity is just that of the noiseless channel: $\log d$
(in this paper $\log$ and $\exp$ are to basis $2$).
\par
However, for entangled $\rho$ the phenomenon of \emph{dense coding}
arises, which was first described in~\cite{bennett:wiesner}:
there $d_A=d_B=d=2$ was considered, with the joint singlett state
$$\rho=\ketbra{\Psi^-}=\frac{1}{2}\big(\ket{01}-\ket{10}\big)\!
                                  \big(\bra{01}-\bra{10}\big).$$
It was demonstrated that by applying one of the three Pauli
unitaries $\sigma_x,\sigma_y,\sigma_z$, or the identity $\1$,
A can drive the state to any of the four Bell states,
hence can encode $2$ bits. It is quite clear that by starting with
any maximally entangled state, e.g.
$$\rho=\ketbra{\Phi},\text{ with }
                 \ket{\Phi}=\frac{1}{\sqrt{d}}\sum_{i=0}^{d-1} \ket{i}\otimes\ket{i}$$
on the system $\C^d\otimes\C^d$, i.e. $d_A=d_B=d$,
one can devise a scheme to transmit $2\log d=\log d^2$ bits
(see~\cite{werner:tele:denseco} for a detailed discussion).
\par
It is less clear what happens if the state is not maximally entangled, or
even mixed: however, since the protocol A and B have to follow
depends even in the maximally mixed case on the actual state, we allow
them to use \emph{the protocol optimally adapted to $\rho$}. Formally:
\par
A chooses an operation (i.e. a completely positive, trace preserving
linear map)
$$T:\alg{L}({\cal H}_A)\longrightarrow\alg{L}({\cal H}),$$
and applies it to her part of $\rho$, after which she sends the resulting
state to B. He thus receives the signal state
$$\rho^T:=(T\otimes\id)\rho.$$
We here assume that one copy of $\rho$ is available per use of the noiseless
channel. Below we will discuss the case of more or unlimited many
copies per round.
\par
Then we can compute the mutual information
$$I(\mu;\rho):=H\left(\int {\rm d}\mu(T)\rho^T\right)
                     -\int {\rm d}\mu(T)H\left(\rho^T\right),$$
with respect to a probability measure $\mu$ on the space
${\bf CP}({\cal H}_A,{\cal H})$
of quantum operations (i.e., completely positive, trace preserving,
linear maps) from ${\cal B}({\cal H}_A)$ to ${\cal B}({\cal H})$.
By the quantum channel coding theorem, eq.~(\ref{eq:capacity}),
of~\cite{holevo:coding} and~\cite{sw:coding}, the \emph{dense coding capacity}
$$DC(d,\rho):=\sup_\mu I(\mu;\rho)$$
is the classical capacity of the channel with signal states $\rho^T$,
when block coding using product states
\begin{equation*}\begin{split}
  \rho^{T_1}\otimes\cdots\otimes\rho^{T_n}
    &=\big((T_1\otimes\id)\rho\big)\otimes\cdots\otimes\big((T_n\otimes\id)\rho\big) \\
    &=\left((T_1\otimes\cdots\otimes T_n)\otimes\id^{\otimes n}\right)\rho^{\otimes n}
\end{split}\end{equation*}
is allowed.
\par
If we impose no restriction on the block coding, i.e. all states
$$({\bf T}\otimes\id^{\otimes n})\rho^{\otimes n},$$
with ${\bf T}\in{\bf CP}({\cal H}_A^{\otimes n},{\cal H}^{\otimes n})$
are admissible, we get the \emph{ultimate dense coding capacity}
$$\overline{DC}(d,\rho)
     =\lim_{n\rightarrow\infty} \frac{1}{n}DC\left(d^n,\rho^{\otimes n}\right).$$
Note that the limit exists by the trivial superadditivity of $DC$:
$$DC(d_1d_2,\rho\otimes\sigma)\geq DC(d_1,\rho)+DC(d_2,\sigma).$$
\par
Our first task is the evaluation of $DC(d,\rho)$:
\par
Assume any probability distribution $\mu$ on ${\bf CP}({\cal H}_A,{\cal H})$,
and denote the Haar measure on the group ${\cal U}(d)$ of unitaries
on ${\cal H}$ as ${\rm d}U$. Then for
every unitary $U$ we have (by unitary invariance of entropy)
\begin{align*}
  H\left(\int {\rm d}\mu(T)\rho^T\right) 
             &=H\left(\int {\rm d}\mu(T)(U\otimes\1)\rho^T(U\otimes\1)^*\right), \\
  H\left(\rho^T\right)
             &=H\left((U\otimes\1)\rho^T(U\otimes\1)^*\right).
\end{align*}
I.e. $I(\mu;\rho)=I(\mu^U;\rho)$, with the translated measure
$$\mu^U(F)=\mu(U^* F U),\text{ for measureable }F\subset{\bf CP}({\cal H}_A,{\cal H}).$$
With concavity of $H$ we find
\begin{equation*}\begin{split}
  I(\mu;\rho) &=    \int_{{\cal U}(d)} \!\!{\rm d}U\left[
                    H\left(\int {\rm d}\mu(T)(U\otimes\1)\rho^T(U\otimes\1)^*\right)
                                                                           \right. \\
              &\phantom{=======}-\int {\rm d}\mu(T)
                     H\left((U\otimes\1)\rho^T(U\otimes\1)^*\right)\!\biggr]          \\
              &\leq H\left(\int {\rm d}\mu(T)\int {\rm d}U
                                           (U\otimes\1)\rho^T(U\otimes\1)^*\right) \\
              &\phantom{================}-\int {\rm d}\mu(T) H\left(\rho^T\right).
\end{split}\end{equation*}
The latter quantity is exactly $I(\overline{\mu};\rho)$, with
$\overline{\mu}=\int {\rm d}U \mu^U$.
\par
Now it is straightforward to prove (essentially
by Schur's lemma) that
$$\int {\rm d}U (U\otimes\1)\rho^T(U\otimes\1)^*=\frac{1}{d}\1\otimes\rho_B,$$
where we observed that by definition
$$\tr_A\rho^T=\tr_A\big((T\otimes\id)\rho\big)=\tr_A\rho=\rho_B.$$
Hence maximization yields
$$DC(d,\rho)=\log d+H(\rho_B)-\inf_\mu \int {\rm d}\mu(T) H\left(\rho^T\right).$$
This infimum in turn is achieved at the point mass on a $T$ minimizing
$H\left(\rho^T\right)$.
\par
Hence we arrive at the result
\begin{satz}
  \label{satz:D}
  The dense coding capacity of the state $\rho$ and
  a $d$--level noiseless transmission system, using one copy of $\rho$
  per round and product states for coding, is given by
  $$DC(d,\rho)=\log d+H(\rho_B)-\min_{T} H\big((T\otimes\id)\rho\big),$$
  where the minimization is over all quantum operations
  $T:{\cal B}({\cal H}_A)\rightarrow {\cal B}({\cal H})$.
  \qed
\end{satz}
As a consequence we get:
\begin{satz}
  \label{satz:Dbar}
  Without the restriction on product state encoding, but
  still using one copy of $\rho$ per round, the capacity is
  $$\overline{DC}(d,\rho) \!=\! \log d+H(\rho_B)
                               -\!\lim_{n\rightarrow\infty}\!\frac{1}{n}\min_{{\bf T}}
                                H\!\left(({\bf T}\otimes\id^{\otimes n})\rho^{\otimes n}\right) \!,$$
  where the minimization is over all quantum operations
  ${\bf T}:{\cal B}({\cal H}_A^{\otimes n})\rightarrow {\cal B}({\cal H}^{\otimes n})$.
  \qed
\end{satz}
Note that the argument describes at the same time a distribution
on ${\bf CP}({\cal H}_A,{\cal H})$ that achieves the capacity: A should
apply a fixed minimizing $T$, followed by uniformly distributed
unitary rotations. The effect of the latter can be achieved equally
by a uniform distribution on an orthogonal basis of unitaries
(with respect to the Hilbert--Schmidt inner product $(A,B)=\tr A^*B$
on operators), see~\cite{werner:tele:denseco}.
\par
As applications of the theorem we can see immediately that
for pure states $\ket{\psi}$
$$DC(d,\ketbra{\psi})=\log d+E(\psi)=\log d+H(\tr_B\ketbra{\psi}),$$
a result already reported in~\cite{barenco:ekert}
and~\cite{hausladen:et:al}, and that $DC(d,\rho)=\log d$
if $\rho$ is separable (below, theorem~\ref{satz:D:RE}, we will
see that this holds true even for \emph{non--distillable}
$\rho$): in the first case the optimizing $T$ is any unitary map,
in the second case it is the projection onto any pure
state (note that $DC(d,\rho)\leq\log d$ follows
from the inequality $H(\sigma_B)-H(\sigma)\leq 0$ for
separable $\sigma$).
This latter choice shows that always $DC(d,\rho)\geq\log d$
(it amounts to ignoring the correlation provided by $\rho$).
\par
In general, however, the minimization required by the theorem
seems not an easy task itself. 
\begin{rem}
  \label{rem:coherent}
  The quantity $H(\sigma_B)-H(\sigma)$,
  $\sigma=(T\otimes\1)\rho$, from theorem~\ref{satz:D} has appeared in another
  context before: it is the \emph{coherent} information of
  Schumacher~\cite{schumacher:coherent}.
\end{rem}
\begin{rem}
  \label{rem:hhhlt}
  Until now we stuck to using one copy of $\rho$ per use of the
  noiseless channel. In recent work by Horodecki et al.~\cite{hhhlt:ndc}
  this restriction was lifted: unlimited many copies of $\rho$ were
  assumed to be available.
  Of course, the theorem can be used to obtain a formula for that case,
  too, which we give, because it interestingly differs from
  than the one in~\cite{hhhlt:ndc} (though of course the numbers coincide):
  \par
  Assume $k$ copies of $\rho$ may be used per round. Obviously
  the resulting dense coding capacity is
  $${DC}^{(k)}(d,\rho)=DC(d,\rho^{\otimes k}),$$
  and for unlimited use of $\rho$ we get
  $${DC}^{(\infty)}(d,\rho)=\lim_{k\rightarrow\infty} DC(d,\rho^{\otimes k}).$$
  Similarly for the ultimate dense coding capacity with $k$ copies
  of $\rho$ per round:
  \begin{equation*}\begin{split}
    \overline{{DC}^{(k)}}(d,\rho) &=\overline{DC}(d,\rho^{\otimes k})                \\
                                  &=\lim_{n\rightarrow\infty}\frac{1}{n} DC(d^n,\rho^{\otimes kn}),
  \end{split}\end{equation*}
  and with unlimited use of $\rho$:
  \begin{equation*}\begin{split}
    \overline{{DC}^{(\infty)}}(d,\rho) &=\lim_{k\rightarrow\infty} \overline{{DC}^{(k)}}(d,\rho)     \\
                                  &=\lim_{k\rightarrow\infty}
                                     \lim_{n\rightarrow\infty} \frac{1}{n} DC(d^n,\rho^{\otimes kn}) \\
                                  &=\lim_{n\rightarrow\infty}
                                     \lim_{k\rightarrow\infty} \frac{1}{n} DC(d^n,\rho^{\otimes kn}) \\
                                  &=\lim_{n\rightarrow\infty} \frac{1}{n} {DC}^{(\infty)}(d,\rho)    \\
                                  &=\overline{DC}^{(\infty)}(d,\rho).
  \end{split}\end{equation*}
  (The limits are exchangeable because the double $\lim$ is actually a joint $\sup$
  over $n$ and $k$, because of monotonicity).
  \par
  In~\cite{hhhlt:ndc} the differently looking expression (for the case $d=2$)
  $$\overline{DC}(\rho)=\sup_n \sup_{\bf T} \! \left\{
      \! 1 \!+\! \frac{nH(\rho_B) \!-\! H\left(({\bf T}\otimes\id^{\otimes n})(\rho^{\otimes n})\right)}
             {H\left({\bf T}(\rho_A^{\otimes n})\right)} \! \right\} \!,$$
  was given, the $\sup$ being over all quantum operations ${\bf T}$
  defined on ${\cal B}({\cal H}_A^{\otimes n})$. However, the derivation in that
  work is sufficiently close to ours so as see identity of the results.
\end{rem}
\par
Let us comment here a bit on other related work, and the relation
of $DC(d,\rho)$ to entanglement:
\par
In the works~\cite{bose:plenio:vedral} and~\cite{bowen}
the relation of the dense coding capacity
to entanglement measures was stressed.
With our results, it is easy to reproduce the observations
of these papers, and go even a little further:
\par
We use the following inequality from~\cite{plenio:virmani:papadopoulos}:
for a (two--way) non--distillable state $\sigma$
$$H(\rho_B)-H(\rho)\leq D(\rho\|\sigma).$$
Applying $T\otimes\id$ to both $\rho$ and $\sigma$, and
invoking the monotonicity of relative entropy under completely
positive maps~\cite{lindblad,uhlmann} we find
\begin{equation*}\begin{split}
  H(\rho_B)-H\big((T\otimes\id)\rho\big)
         &\leq D\big((T\otimes\id)\rho\|(T\otimes\id)\sigma\big) \\
         &\leq D(\rho\|\sigma).
\end{split}\end{equation*}
Now minimize over $T$ and non--distillable $\sigma$: this proves
\begin{satz}
  \label{satz:D:RE}
  For all states $\rho$ one has
  $$DC(d,\rho)\leq \log d+E_{\rm re}(\rho),$$
  where $E_{\rm re}(\rho)=\inf_{\sigma\in{\cal D}} D(\rho\|\sigma)$
  is the relative entropy of entanglement with respect
  to the set ${\cal D}$ of non--distillable states.
  \qed
\end{satz}
In particular, when $\rho$ is non--distillable, $DC(d,\rho)=\log d$
(see also~\cite{hhhlt:ndc} for this observation). One may
wonder, whether the inverse is true, too: when $\rho$ is
distillable, does $DC(d,\rho)>\log d$ follow?
\par
To compare this result to the statements in~\cite{bose:plenio:vedral,bowen},
and the result of the recently published~\cite{hiroshima:dense}
we have to note that in these works only \emph{unitary} encodings
were considered.
Hence our $DC(d,\rho)$ is typically a strict
upper bound to the capacity in the cited works.
Still, questions raised in~\cite{bose:plenio:vedral,bowen}
receive answers: the conjectured capacity formulas and inequalities
from these works follow immediately, by the same method of
Haar averaging we employed above (see also~\cite{hiroshima:dense}).
\par
To get a bound in the other direction is not so easy.
We might try to go further on the road of entanglement,
and find an entanglement measure lower bound. For example,
if we could prove that
$$f(\rho)=\overline{DC}^{(\infty)}(d,\rho)-\log d$$
is an entanglement measure itself, we would find the lower bound
$$\overline{DC}^{(\infty)}(d,\rho)\geq \log d+E_D(\rho),$$
with the \emph{distillable entanglement} $E_D(\rho)$:
this follows from general inequalities in~\cite{horodecki:ent:measures}.
We leave this question, however, to another occasion.
\par\medskip
We would like now to discuss the additivity of $D$, i.e. whether for
states $\rho$, $\sigma$
\begin{equation}
  \label{eq:add}
  DC(d_1d_2,\rho\otimes\sigma)=DC(d_1,\rho)+DC(d_2,\sigma).
\end{equation}
Note that if this is true for $\rho$ and all $\sigma=\rho^{\otimes n}$ (e.g.,
inductively), it immediately follows that
$\overline{DC}(d,\rho)=DC(d,\rho)$. In particular, all ultimate capacities
in remark~\ref{rem:hhhlt} are identical to their ``un--barred''
versions. The capacity with unlimited use of $\rho$ from~\cite{hhhlt:ndc}
would then read
\begin{equation*}\begin{split}
  \overline{DC}&^{(\infty)}(d,\rho) ={DC}^{(\infty)}(d,\rho)                     \\
               &\phantom{===}=\log d+H(\rho_B)
                              -\inf_k\min_T H\bigl((T\otimes\id^{\otimes k})\rho^{\otimes k}\bigr),
\end{split}\end{equation*}
where the minimization is over all quantum operations
$T:{\cal B}({\cal H}_A^{\otimes k})\rightarrow {\cal B}({\cal H})$.
\par
By theorem~\ref{satz:D}, the statement of
eq.~(\ref{eq:add}) is equivalent to asking, if
\begin{equation*}\begin{split}
  \min_{T_{12}} H\big((T_{12}\otimes\id^{\otimes 2})(\rho\otimes\sigma)\big)
     &=\min_{T_1} H\big((T_1\otimes\id)\rho\big)              \\
     &\phantom{==}+\min_{T_2} H\big((T_2\otimes\id)\sigma\big).
\end{split}\end{equation*}
Obviously, and fitting with the superadditivity of $D$,
``$\leq$'' (subadditivity) is trivial, and the question is if
``$<$'' can occur. Note that in this generality it is quite easy
to come up with states that violate the additivity property,
see the discussion below. The problem is rather to find conditions
where additivity holds.
%%%%%%%%%%%%%%%%%%%%%%%%%%%%%%%%%%%%%%%%%%%%%%%%%%%%%%%%%%%%%%%%%%%%%%%%%%%%%%%%%%%%
%
%\par
%... ??? ... One possible idea is to consider purifications of
%the states, and Stinespring dilations of the operations, to
%have them unitary. Maybe it can then be shown that it is always
%unnecessary to entangle the two initially separate systems (???).
%
%%%%%%%%%%%%%%%%%%%%%%%%%%%%%%%%%%%%%%%%%%%%%%%%%%%%%%%%%%%%%%%%%%%%%%%%%%%%%%%%%%%%
\par\medskip
Generalizing, one may assume not a noiseless, but a noisy channel
$\varphi:{\cal B}({\cal H})\rightarrow {\cal B}({\cal H})$
between A and B, and consider the dense coding capacities
\begin{align*}
  DC(\varphi,\rho),              & \quad \overline{DC}(\varphi,\rho),           \\
  {DC}^{(k)}(\varphi,\rho),      & \quad \overline{DC}^{(k)}(\varphi,\rho),     \\
  {DC}^{(\infty)}(\varphi,\rho), & \quad \overline{DC}^{(\infty)}(\varphi,\rho).
\end{align*}
For example, we can define
$$DC(\varphi,\rho)=\sup_\mu I(\mu;\varphi\circ\rho),$$
over all probability distributions $\mu$ on
${\bf CP}({\cal H}_A,{\cal H})$, with
$$I(\mu;\varphi\circ\rho):=H\left(\int {\rm d}\mu(T)\rho^{\varphi\circ T}\right)
                                 -\int {\rm d}\mu(T)H\left(\rho^{\varphi\circ T}\right).$$
Observe that our previous $DC(d,\rho)$ is reproduced in the new
definition as $DC(\id_d,\rho)$. Further, observe that for a pure entangled
state $\rho$ the definition relates to the entanglement assisted
classical capacity~\cite{ent:ass:class:cap} of the quantum channel
$\varphi$: in fact, $\overline{DC}^{(\infty)}(\varphi,\rho)$
\emph{is} this latter quantity.
\par
Again, the superadditivity
\begin{equation}
  \label{eq:gen:add}
  DC(\varphi\otimes\vartheta,\rho\otimes\sigma)
                      \geq DC(\varphi,\rho)+DC(\vartheta,\sigma)
\end{equation}
trivially holds, and we may study conditions for equality
in eq.~(\ref{eq:gen:add}), i.e. additivity.
\par
Note that it is fairly easy to come up with situations ($\varphi$, $\vartheta$,
$\rho$, $\sigma$) where strict superadditivity holds. In fact one can even
have either $\varphi=\vartheta$ or $\rho=\sigma$: e.g.~consider
$$\varphi=\vartheta=\id_{{\cal B}(\C^2)},\quad
\left\{\begin{array}{l}
         \rho=\ketbra{00}\text{ (unentangled)}, \\
         \sigma=\ketbra{\Psi^-}^{\otimes 2},
       \end{array}\right.$$
or alternatively
$$\left.\begin{array}{l}
          \varphi=\id_{{\cal B}(\C^4)}, \\
          \vartheta=\frac{1}{2}\1\text{ (constant map)},
        \end{array}\right\}
\quad \rho=\sigma=\ketbra{\Psi^-}.$$
But with both these conditions simultaneously it seems not so easy.
It may even be that \emph{weak additivity} holds, i.e.
$$DC(\varphi^{\otimes n},\rho^{\otimes n})=n\, DC(\varphi,\rho),$$
for all channels $\varphi$ and joint states $\rho$, but we could
not reach a conclusive result on this question.

\section{Reductions among\protect\\ additivity questions}
\label{sec:reductions}
We have encountered two paradigms of coding in quantum channels,
the first in the established discussion (a good overview is
in~\cite{holevo:overview}, and some recent developments are
reviewed in~\cite{ahw:add}), the second in the previous section:
\par\medskip\noindent
{\bf 1. State preparation}: {\it The encoder may prepare any
state on the input system space ${\cal H}_1$ for the quantum channel
$\varphi:{\cal B}({\cal H}_1)\rightarrow{\cal B}({\cal H}_2)$.}
\par\medskip\noindent
{\bf 2. Action on given state}: {\it On the input system a
state is given in advance (possibly entangled with the receiver),
and the encoder may act on it in an arbitary way, and the result
is sent down the channel $\varphi$.}
\par\medskip
It is quite obvious that method 1 can be reduced to method 2:
the previously given state is just any state not entangled with
the receiver (say, a pure state). Then by executing an appropriate operation
the encoder can drive the input into any desired state.
\par
Less obvious, but still quite canonical, is the converse reduction:
any operation $T:{\cal B}({\cal H}_A)\rightarrow{\cal B}({\cal H}_1)$
can be implemented as a unitary
$$U:{\cal H}_A\otimes{\cal H}' \longrightarrow {\cal H}_1\otimes{\cal H}'',$$
followed by a partial trace over ${\cal H}''$, the system
${\cal H}'$ being prepared initially in a null state $\sigma_0$.
This is a formulation of the Stinespring dilation theorem~\cite{stinespring},
and it is quite easy to see that $\dim{\cal H}'$ can be chosen fixed
and finite for all possible $T$. Now pick an $\epsilon$--approximating
quantum gate $G_\epsilon$ for ${\cal H}_A\otimes{\cal H}'$, with program
register ${\cal H}_{P_\epsilon}$: by choosing $\psi$ in the
program register appropriately one obtains (using monotonicity
of the trace norm under partial trace), for all states $\sigma$
on ${\cal H}_A$,
\begin{equation}
  \label{eq:red:21}
  \left\| T(\sigma)-\tr_{{\cal H}''\otimes{\cal H}_{P_\epsilon}}
                       \bigl( G_\epsilon(\sigma\otimes\ketbra{0}\otimes\ketbra{\psi})G_\epsilon^* \bigr)
  \right\|_1\leq \epsilon.
\end{equation}
Thus every coding process by acting on the input system can be arbitrarily
well approximated by coding via choice of $\ket{\psi}\in{\cal H}_{P_\epsilon}$.
\par
These two reductions, however, are of a very different nature, as
we can see by considering their behaviour under tensor products
of channels: while the reduction $1\rightarrow 2$ scales alright
(any entangled input state can be obtained by a suitable entangling 
operation on the product of the initial states), the reduction
$2\rightarrow 1$ that we proposed does not.
In fact, as we have seen in theorem~\ref{satz:no:spqg},
on a product ${\cal H}_{A1}\otimes{\cal H}_{A2}$ of two input systems we 
can never implement an entangling operation, once we have chosen
approximating quantum gates for each of them individually according
to eq.~(\ref{eq:red:21}), and tensor them.
\par\medskip
We have seen that there are channels where classical information is
encoded after method 1 (these are just the operations $\varphi$),
and that there are channels where it is encoded after method 2
(the generalized noisy dense coding channels).
The above reductions show that the two approaches are equivalent
in the sense that a channel of the one kind can be simulated
to arbitary accuracy by one of the other kind.
\par
However, for the additivity question of channel capacity one has to
look at higher tensor products of the channel at hand. By the
above argument the reduction $1\rightarrow 2$ provides a reduction
of the additivity question for channels of the first type to those
of the second type. It is unknown to us
if the additivity question can be reduced in the other direction:
the construction above, summarized in eq.~(\ref{eq:red:21}),
at least does not provide this, as we have seen.
On the other hand, it appears to be most natural: it seems the
most reasonable thing to do
to associate a channel of the first type to the given channel of
the second type that has the same properties with respect to
classical information transmission, by simply enabling to
emulate the effect of any encoding transformation $T$
by a suitable input state.

\section{Conclusion}
\label{sec:conclusion}
By studying entanglement assisted classical communication via
quantum channels, attention was drawn towards channels which
require \emph{actions} for signalling rather than \emph{state preparations}
like the usual quantum dynamics, represented by completely positive maps.
An attempted reduction of the more general scenario to the usual
one was shown to fail, because no \emph{scalable programmable quantum gates}
exist. This was taken to indicate that the new concept is strictly
more general, which leads us to conjecture that the additivity
question for quantum channel capacity really is not about ``whether entangled
inputs help'', but rather ``whether \emph{entangling} inputs help''.
It must be stressed that in the more general vista we presented,
additivity is not a mre matter of ``right'' or ``false''. Rather
it becomes (as we demonstrated by examples) a question of
\emph{characterization} of the situations where it holds.
Note finally that the very occurence of the above mentioned
distinction in coding concepts is a purely quantum phenomenon.

\acknowledgments
Part of this work was carried out during the author's visits
to Universit\"at Bielefeld (July 2001), and the JST ERATO project
``Quantum Computation and Information'', Tokyo (August/September 2001).
The hospitality of these institutions is gratefully acknowledged.

% bibliographie

%\bibliographystyle{revtex}

\end{document}